\documentclass{ws-ijmpcs}

\begin{document}

\newcommand{\be}{\begin{equation}}
\newcommand{\ee}{\ear\noindent}
\newcommand{\bear}{\begin{eqnarray}}
\newcommand{\ear}{\end{eqnarray}\noindent}
\newcommand{\no}{\noindent}
\newcommand{\veject}{\vfill\eject}
\newcommand{\ven}{\vfill\eject\noindent}
\date{}
\renewcommand{\theequation}{\arabic{section}.\arabic{equation}}
\renewcommand{\arraystretch}{2.5}
\newcommand{\GeV}{\mbox{GeV}}
\newcommand{\cL}{\cal L}
\newcommand{\D}{\cal D}
\newcommand{\Dhalf}{{D\over 2}}
\newcommand{\Det}{{\rm Det}}
\newcommand{\PP}{\cal P}
\newcommand{\G}{{\cal G}}
\def\R{1\!\!{\rm R}}
\def\Eins{\mathord{1\hskip -1.5pt
\vrule width .5pt height 7.75pt depth -.2pt \hskip -1.2pt
\vrule width 2.5pt height .3pt depth -.05pt \hskip 1.5pt}}
\newcommand{\symb}{\mbox{symb}}
\renewcommand{\arraystretch}{2.5}
\newcommand{\slD}{\raise.15ex\hbox{$/$}\kern-.57em\hbox{$D$}}
\newcommand{\slpartial}{\raise.15ex\hbox{$/$}\kern-.57em\hbox{$\partial$}}
\newcommand{\slG}{{{\dot G}\!\!\!\! \raise.15ex\hbox {/}}}
\newcommand{\Gd}{{\dot G}}
\newcommand{\Gund}{{\underline{\dot G}}}
\newcommand{\Gdd}{{\ddot G}}
\def\slash#1{#1\!\!\!\raise.15ex\hbox {/}}
\def\GBd12{{\dot G}_{B12}}
\def\Gphat#1#2{{\hat{{\dot {{\cal G}}}}_{B#1#2}}}
\def\mneg{\!\!\!\!\!\!\!\!\!\!}
\def\Mneg{\!\!\!\!\!\!\!\!\!\!\!\!\!\!\!\!\!\!\!\!}
\def\exmn{\Bigl(\mu \leftrightarrow \nu \Bigr)}
\def\non{\nonumber}
\def\beqn*{\begin{eqnarray*}}
\def\eqn*{\end{eqnarray*}}
\def\sy{\scriptscriptstyle}
\def\footstrut{\baselineskip 12pt}
\def\square{\kern1pt\vbox{\hrule height 1.2pt\hbox{\vrule width 1.2pt
   \hskip 3pt\vbox{\vskip 6pt}\hskip 3pt\vrule width 0.6pt}
   \hrule height 0.6pt}\kern1pt}
\def\slash#1{#1\!\!\!\raise.15ex\hbox {/}}
\def\dint#1{\int\!\!\!\!\!\int\limits_{\!\!#1}}
\def\bra#1{\langle #1 |}
\def\ket#1{| #1 \rangle}
\def\vev#1{\langle #1 \rangle}
\def\rightvac{\mid 0\rangle}
\def\leftvac{\langle 0\mid}
\def\dps{\displaystyle}
\def\sy{\scriptscriptstyle}
\def\half{{1\over 2}}
\def\third{{1\over3}}
\def\fourth{{1\over4}}
\def\fifth{{1\over5}}
\def\sixth{{1\over6}}
\def\seventh{{1\over7}}
\def\eigth{{1\over8}}
\def\ninth{{1\over9}}
\def\tenth{{1\over10}}
\def\pa{\partial}
\def\ddtau{{d\over d\tau}}
\def\ge{\hbox{\textfont1=\tame $\gamma_1$}}
\def\gz{\hbox{\textfont1=\tame $\gamma_2$}}
\def\gd{\hbox{\textfont1=\tame $\gamma_3$}}
\def\go{\hbox{\textfont1=\tamt $\gamma_1$}}
\def\gt{\hbox{\textfont1=\tamt $\gamma_2$}}
\def\gth{\hbox{\textfont1=\tamt $\gamma_3$}} 
\def\gf{\hbox{$\gamma_5\;$}}
\def\ie{\hbox{$\textstyle{\int_1}$}}
\def\iz{\hbox{$\textstyle{\int_2}$}}
\def\id{\hbox{$\textstyle{\int_3}$}}
\def\ldop{\hbox{$\lbrace\mskip -4.5mu\mid$}}
\def\rdop{\hbox{$\mid\mskip -4.3mu\rbrace$}}
\def\eps{\epsilon}
\def\epshalf{{\epsilon\over 2}}
\def\e{\mbox{e}}
\def\g{\mbox{g}}
\def\kinb{{1\over 4}\dot x^2}
\def\kinf{{1\over 2}\psi\dot\psi}
\def\expk{{\rm exp}\biggl[\,\sum_{i<j=1}^4 G_{Bij}k_i\cdot k_j\biggr]}
\def\expp{{\rm exp}\biggl[\,\sum_{i<j=1}^4 G_{Bij}p_i\cdot p_j\biggr]}
\def\expshort{{\e}^{\half G_{Bij}k_i\cdot k_j}}
\def\expabb{{\e}^{(\cdot )}}
\def\epseps#1#2{\varepsilon_{#1}\cdot \varepsilon_{#2}}
\def\epsk#1#2{\varepsilon_{#1}\cdot k_{#2}}
\def\kk#1#2{k_{#1}\cdot k_{#2}}
\def\G#1#2{G_{B#1#2}}
\def\Gp#1#2{{\dot G_{B#1#2}}}
\def\GF#1#2{G_{F#1#2}}
\def\Dab{{(x_a-x_b)}}
\def\Dsq{{({(x_a-x_b)}^2)}}
\def\lag{( -\partial^2 + V)}
\def\PITD{{(4\pi T)}^{-{D\over 2}}}
\def\4piTD{{(4\pi T)}^{-{D\over 2}}}
\def\4piT4{{(4\pi T)}^{-2}}
\def\TintmD{{\dps\int_{0}^{\infty}}\frac{dT}{T}\,e^{-m^2T}
    {(4\pi T)}^{-{D\over 2}}}
\def\go{\hbox{\textfont1=\tamt $\gamma_1$}}
\def\gt{\hbox{\textfont1=\tamt $\gamma_2$}}
\def\gth{\hbox{\textfont1=\tamt $\gamma_3$}} 
\def\gf{\hbox{$\gamma_5\;$}}
\def\ie{\hbox{$\textstyle{\int_1}$}}
\def\iz{\hbox{$\textstyle{\int_2}$}}
\def\id{\hbox{$\textstyle{\int_3}$}}
\def\ldop{\hbox{$\lbrace\mskip -4.5mu\mid$}}
\def\rdop{\hbox{$\mid\mskip -4.3mu\rbrace$}}
\def\eps{\epsilon}
\def\epshalf{{\epsilon\over 2}}
\def\e{\mbox{e}}
\def\g{\mbox{g}}
\def\kinb{{1\over 4}\dot x^2}
\def\kinf{{1\over 2}\psi\dot\psi}
\def\expk{{\rm exp}\biggl[\,\sum_{i<j=1}^4 G_{Bij}k_i\cdot k_j\biggr]}
\def\expp{{\rm exp}\biggl[\,\sum_{i<j=1}^4 G_{Bij}p_i\cdot p_j\biggr]}
\def\expshort{{\e}^{\half G_{Bij}k_i\cdot k_j}}
\def\expabb{{\e}^{(\cdot )}}
\def\epseps#1#2{\varepsilon_{#1}\cdot \varepsilon_{#2}}
\def\epsk#1#2{\varepsilon_{#1}\cdot k_{#2}}
\def\kk#1#2{k_{#1}\cdot k_{#2}}
\def\G#1#2{G_{B#1#2}}
\def\Gp#1#2{{\dot G_{B#1#2}}}
\def\GF#1#2{G_{F#1#2}}
\def\Dab{{(x_a-x_b)}}
\def\Dsq{{({(x_a-x_b)}^2)}}
\def\lag{( -\partial^2 + V)}
\def\PITD{{(4\pi T)}^{-{D\over 2}}}
\def\4piTD{{(4\pi T)}^{-{D\over 2}}}
\def\4piT4{{(4\pi T)}^{-2}}
\def\TintmD{{\dps\int_{0}^{\infty}}\frac{dT}{T}\,e^{-m^2T}
    {(4\pi T)}^{-{D\over 2}}}
\def\Tintm4{{\dps\int_{0}^{\infty}}\frac{dT}{T}\,e^{-m^2T}
    {(4\pi T)}^{-2}}
\def\Tintm{{\dps\int_{0}^{\infty}}\frac{dT}{T}\,e^{-m^2T}}
\def\Tint{{\dps\int_{0}^{\infty}}\frac{dT}{T}}
\def\pint{{\dps\int}{dp_i\over {(2\pi)}^d}}
\def\Dx{\dps\int{\cal D}x}
\def\Dy{\dps\int{\cal D}y}
\def\Dpsi{\dps\int{\cal D}\psi}
\def\Tr{{\rm Tr}\,}
\def\tr{{\rm tr}\,}
\def\sumij{\sum_{i<j}}
\def\freeexp{{\rm e}^{-\int_0^Td\tau {1\over 4}\dot x^2}}
\def\arraystretch{2.5}
\def\Ge{\mbox{GeV}}
\def\dA{\partial^2}
\def\DA{\sqsubset\!\!\!\!\sqsupset}
\def\FFdual{F\cdot\tilde F}
\def\mn{\mu\nu}

\markboth{Authors' Names}
{}

%
\catchline{}{}{}{}{}
%

\title{\bf THE EULER-HEISENBERG LAGRANGIAN BEYOND ONE LOOP
}

\author{IDRISH HUET
}

\address{Theoretisch-Physikalisches Institut, Friedrich-Schiller-Universit\"at Jena,\\
Max-Wien-Platz 1, D-07743 Jena, Germany\\
idrish.huet@uni-jena.de}

\author{MICHEL RAUSCH DE TRAUBENBERG}

\address{IPHC-DRS, UdS, IN2P3,\\
23 rue du Loess, F-67037 Strasbourg Cedex, France\\
Michel.Rausch@IReS.in2p3.fr}

\author{CHRISTIAN SCHUBERT}

\address{Instituto de F{{\'\i}}sica y Matem\'aticas, Universidad Michoacana de San Nicol\'as de Hidalgo\\
Apdo. Postal 2-82, C.P. 58040, Morelia, Michoacan, Mexico\\
schubert@ifm.umich.mx}

\maketitle

\begin{history}
\received{Day Month Year}
\revised{Day Month Year}
\end{history}

\begin{abstract}
We review what is presently known about higher loop corrections to the
Euler-Heisenberg Lagrangian and its Scalar QED analogue. The use of
those corrections as a tool for the study of the properties of the QED perturbation
series is outlined. As a further step in a long-term effort to prove or disprove
the convergence of the $N$ photon amplitudes in the quenched approximation,
we present a parameter integral representation of the three-loop Euler-Heisenberg Lagrangian
in 1+1 dimensional QED, obtained in the worldline formalism. 

\end{abstract}

\ccode{PACS numbers: 11.10.Kk,12.20.Ds.}

\section{Introduction}
\renewcommand{\theequation}{1.\arabic{equation}}
\setcounter{equation}{0}

Heisenberg and Euler's 1936 calculation \cite{eulhei} of the one-loop effective
Lagrangian induced for a constant Maxwell field by a spinor loop
was not only a milestone in the development of QED, but remains until today the
prototypical example for the concept of integrating out degrees of freedom
in field theory. In practical terms, it encodes in a very concise form information
on a host of photonic low-energy processes (see \refcite{dunneeulhei} for a review).
Higher loop corrections to the Euler-Heisenberg Lagrangian and its Scalar QED analogue,
due to Weisskopf \cite{weisskopf} (both called ``EHL'' in the following for simplicity) 
were studied only much later, starting with Ritus' 1975 calculation \cite{ritusspin} 
of the two-loop EHL.
The purpose of this talk is to give a summary on what is known about the EHL at
the multiloop level, and to argue that those multiloop corrections, although not
likely to be of phenomenological interest in the near future, contain important structural
information on QED. 

\section{QED in a constant external field in the worldline formalism}
\renewcommand{\theequation}{2.\arabic{equation}}
\setcounter{equation}{0}

We start with a short introduction to the worldline representation of the QED S-matrix, going
back to Feynman  \cite{feynman:pr80,feynman:pr84},  
which in recent years has emerged as an extremely efficient
tool for the computation of processes involving constant external fields in QED.
For the simplest case of the one-loop effective action in Scalar QED, it reads
\cite{feynman:pr80}

\begin{equation}
\Gamma\lbrack A\rbrack = \int_0^{\infty}
\frac{dT}{T} \, {\rm e}^{-m^2T}
\int {\cal D}x\, 
{\rm exp} \left[ 
- \int_0^T \!\!\! d\tau \left( {1\over 4}{\dot x}^2 
+ ieA_{\mu}\dot x^{\mu} 
\right) \right]
\label{scalarpi}
\end{equation}
\no
Here $m$ and $T$ are the mass and proper-time of the scalar particle in the loop, and
the path integral runs over the space of closed trajectories
with period $T$, $x^{\mu}(T)=x^{\mu}(0)$, in (euclidean) spacetime.
The spinor QED equivalent of (\ref{scalarpi}) is obtained \cite{feynman:pr84} by the
addition of a global
factor of $-\half$, and the insertion of a {\it spin factor} $S[x,A]$ under
the path integral.
The modern way of writing this spin factor is 
in terms of an additional Grassmann path integral
\cite{fradkin},

\bear
S[x,A]  &=&
\int {\cal D}\psi(\tau)
\,
\exp 
\Biggl\lbrack
-\int_0^Td\tau
\Biggl(
\half \psi\cdot \dot\psi -ie \psi^{\mu}F_{\mn}\psi^{\nu}
\Biggr)
\Biggr\rbrack
\label{spinfactorgrass}
\ear
Here the path integration is over the space of anticommuting
functions antiperiodic in proper-time,
$\psi^{\mu}(\tau_1)\psi^{\nu}(\tau_2) = - \psi^{\nu}(\tau_2)\psi^{\mu}(\tau_1)$,
$\psi^{\mu}(T) = - \psi^{\mu}(0)$.

Presently, three quite different
methods are available for computing such path integrals (see \refcite{mepuva} for a review): 
(i) the ``string-inspired approach'',
based on a perturbative expansion and gaussian path integration 
\cite{polbook,berkos,strassler1,mess1,mereview} 
(ii) the ``worldline
instanton approach'',  using a stationary path approximation \cite{afalma,mewlinst1} and 
(iii) the direct numerical
calculation using Monte Carlo techniques \cite{gielan}. 

All three methods have been applied to the calculation of Euler-Heisenberg Lagrangians.
We will explain here only the ``string-inspired'' method; see \refcite{afalma,mewlinst1,mehumcsc}
for the worldline instanton and \refcite{gielan,giekli} for the worldline Monte Carlo approach.
If we expand the interaction exponential,

\bear
{\rm exp}\Bigl[
-\int_0^Td\tau\, ieA_{\mu}\dot x^{\mu}
\Bigr]
=\sum_{N=0}^{\infty}
{{(-ie)}^N\over N!}
\prod_{i=0}^N
\int_0^Td\tau_i
\biggl[
\dot x^{\mu}(\tau_i)
A_{\mu}(x(\tau_i))
\biggr]
\label{expandint}
\ear\no
the individual terms correspond to Feynman diagrams
describing a fixed number of
interactions of the scalar loop with
the external field, see fig. \ref{1loopexpand}.

\begin{figure}[h]
{
$\hspace{40pt}$
\includegraphics{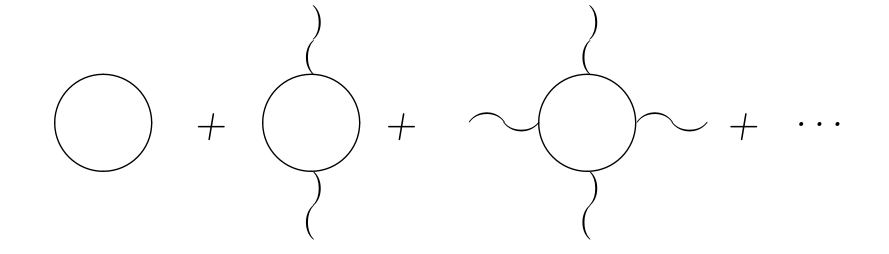}
}
\caption{Feynman diagrams equivalent to the one-loop effective action.}
\label{1loopexpand}
\end{figure}

The corresponding $N$ -- photon
scattering amplitude is then obtained by
specializing to a background
consisting of 
a sum of plane waves with definite
polarizations,
$A_{\mu}(x)=
\sum_{i=1}^N
\varepsilon_{i\mu}
\e^{ik_i\cdot x}
$,
and picking out the term containing every
$\varepsilon_i $ once.
This yields the following representation of
the $N$ - photon amplitude, 

\bear
\Gamma[\lbrace k_i,\varepsilon_i\rbrace\rbrace]
&=&
(-ie)^{N}
\Tintm 
\int {\cal D}x\,
V_{\rm scal}^{A}[k_1,\varepsilon_1]\ldots
V_{\rm scal}^{A}[k_N,\varepsilon_N]
\e^{-\int_0^Td\tau {\dot x^2\over 4}}
\non\\
\label{repNvector}
\ear\no
Here 
$
V_{\rm scal}^{A}[k,\varepsilon]
\equiv
\int_0^Td\tau\,
\varepsilon\cdot \dot x(\tau)
\,{\rm e}^{ikx(\tau)}
$
denotes the same photon vertex operator used in string perturbation theory.
The integral over the zero mode
$x_0^{\mu}= {1\over T}\int_0^T d\tau\, x^{\mu}(\tau)$ factors out
and produces the usual energy-momentum conservation factor.
The reduced path integral $\int{\cal D}y(\tau)$
over $y(\tau)\equiv x(\tau) - x_0$ is gaussian, and
can be evaluated using the ``bosonic''
worldline Green's function $G_B$,

\bear
\langle y^{\mu}(\tau_1)y^{\nu}(\tau_2)\rangle
= -g^{\mu\nu}G_B(\tau_1,\tau_2)=
-g^{\mu\nu}\Bigl[
\mid \tau_1-\tau_2\mid 
-{{(\tau_1-\tau_2)}^2\over T}
\Bigr] 
\label{defGB}
\ear
\no
Using a formal exponentiation of the factors
$\varepsilon_i\cdot\dot x_i$'s and ``completing the
square'' yields the following
closed-form expression for the one-loop
$N$ - photon amplitude:

\begin{eqnarray}
\hspace{-20pt}
\Gamma[\lbrace k_i,\varepsilon_i\rbrace]
&=&
{(-ie)}^N
{(2\pi )}^D\delta (\sum k_i)
{\dps\int_{0}^{\infty}}\frac{dT}{T}
{(4\pi T)}^{-{D\over 2}}
e^{-m^2T}
\prod_{i=1}^N \int_0^T 
d\tau_i
\qquad
\nonumber\\
&&\hspace{-40pt}
\times
\exp\biggl\lbrace\sum_{i,j=1}^N 
\bigl\lbrack \half G_{Bij} k_i\cdot k_j
+i\dot G_{Bij}k_i\cdot\varepsilon_j 
+\half\ddot G_{Bij}\varepsilon_i\cdot\varepsilon_j
\bigr\rbrack\biggr\rbrace
\mid_{\rm multi-linear}
\nonumber\\
\label{scalarqedmaster}
\end{eqnarray}
\no
Here it is understood that only the terms linear
in all the $\varepsilon_1,\ldots,\varepsilon_N$
have to be taken. 
Dots generally denote a
derivative acting on the first variable,
and we abbreviate
$G_{Bij}\equiv G_B(\tau_i,\tau_j)$ etc.
The factor ${(4\pi T)}^{-{D\over 2}}$
represents the free Gaussian path integral
determinant.
The expression (\ref{scalarqedmaster})
is identical with the ``Bern-Kosower
Master Formula'' 
for the $N$ photon case \cite{berkos,strassler1,mereview}. 

In the spinor QED case, the correlator for the evaluation of the
additional Grassmann path integral
is $\langle \psi(\tau_1)\psi(\tau_2)\rangle = 
\half g^{\mu\nu}G_F(\tau_1,\tau_2)$, with 
$G_F(\tau_1,\tau_2) = {\rm sign}(\tau_1-\tau_2)$.
Its explicit evaluation can, however, be circumvented,
using the following ``replacement rule'' \cite{berkos}:
Writing out the exponential in the master formula
eq.(\ref{scalarqedmaster}) for a fixed number $N$ of
photons, one obtains an integrand

\bear \exp\biggl\lbrace 
\biggr\rbrace \mid_{\rm multi-linear} \quad={(-i)}^N P_N(\dot
G_{Bij},\ddot G_{Bij}) \exp\biggl[\half \sum_{i,j=1}^N G_{Bij}k_i\cdot
k_j \biggr] \label{defPN} \ear\no 
with a certain polynomial $P_N$
depending on the various  $\dot G_{Bij}$'s, $\ddot G_{Bij}$'s, 
as well as on the kinematic invariants. 
By suitable partial integrations all second derivatives
$\ddot G_{Bij}$ appearing in $P_N$ can be removed,
so that $P_N$ gets replaced by
another polynomial $Q_N$ depending solely
on the $\dot G_{Bij}$'s,

\bear
P_N(\dot G_{Bij},\ddot G_{Bij})
\,\e^{\half\sum G_{Bij}k_i\cdot k_j}
\quad
{\stackrel{\sy{\rm part. int.}}{\longrightarrow}}
\quad
Q_N(\dot G_{Bij})
\,\e^{\half\sum G_{Bij}k_i\cdot k_j}
\label{partint}
\ear\no
Then the integrand for the spinor loop
case can be obtained by simultaneously
replacing every closed cycle 
$\dot G_{Bi_1i_2}\dot G_{Bi_2i_3}\cdots\dot G_{Bi_ki_1}$
appearing in $Q_N$ by

\begin{equation}
\dot G_{Bi_1i_2}
\dot G_{Bi_2i_3}\cdots\dot G_{Bi_ki_1}
- G_{Fi_1i_2}G_{Fi_2i_3}\cdots
G_{Fi_ki_1}
\label{subrule}
\end{equation}\no

An additional background field $\bar A^{\mu}(x)$ with constant field strength tensor
$\bar F_{\mu\nu}$  can be simply taken into account by appropriate
changes of the worldline propagators and the path integral
determinant. Those are \cite{mess1,shaisultanov,merescsc,mevv} (deleting the ``bar'')

\begin{eqnarray}
G_{B12}&\rightarrow&{\cal G}_{B12}\equiv
{T\over 2{({\cal Z})}^2}\biggl({{\cal Z}\over{{\rm sin}({\cal Z})}}
{\rm e}^{-i{\cal Z}\dot G_{B12}}
+i{\cal Z}\dot G_{B12} -1\biggr)
\nonumber\\
G_{F12}&\rightarrow&{\cal G}_{F12} =
G_{F12}
{{\rm e}^{-i{\cal Z}\dot G_{B12}}\over {\rm cos}({\cal Z})}
\nonumber\\
{(4\pi T)} ^{-{D\over 2}}
&\rightarrow&
{(4\pi T)}^{-{D\over 2}}
{\rm det}^{-{1\over 2}}
\biggl[{\sin({\cal Z})\over {{\cal Z}}}
\biggr] \qquad {(\rm Scalar\quad QED)}
\nonumber\\
{(4\pi T)} ^{-{D\over 2}}
&\rightarrow&
{(4\pi T)}^{-{D\over 2}}
{\rm det}^{-{1\over 2}}
\biggl[{\tan({\cal Z})\over {{\cal Z}}}
\biggr] \qquad {(\rm Spinor\quad QED)}
\nonumber\\
\label{calGBGFdet}
\end{eqnarray}
\noindent
These expressions should be understood as power
series in the matrix ${\cal Z}^{\mu\nu}\equiv eTF^{\mu\nu}$.

Thus one obtains the following generalization of 
(\ref{scalarqedmaster}),
representing the scalar QED $N$ - photon scattering amplitude 
in a constant field \cite{shaisultanov,merescsc}:

\begin{eqnarray}
&&\Gamma_{\rm scal}
[\lbrace k_i,\varepsilon_i\rbrace]
=
{(-ie)}^N
{(2\pi )}^D\delta (\sum k_i)
{\dps\int_{0}^{\infty}}\frac{dT}{T}
{(4\pi T)}^{-{D\over 2}}
e^{-m^2T}
{\rm det}^{-{1\over 2}}
\biggl[{{\rm sin}({\cal Z})\over {\cal Z}}\biggr]
\non\\
&&\hspace{2pt}\times
\prod_{i=1}^N \int_0^T 
d\tau_i
\exp\biggl\lbrace\sum_{i,j=1}^N 
\Bigl\lbrack \half k_i\cdot {\cal G}_{Bij}\cdot  k_j
-i\varepsilon_i\cdot\dot{\cal G}_{Bij}\cdot k_j
+\half
\varepsilon_i\cdot\ddot {\cal G}_{Bij}\cdot\varepsilon_j
\Bigr\rbrack\biggr\rbrace
\mid_{\rm multi-linear}\quad
\nonumber\\
\label{scalarqedmasterF}
\end{eqnarray}
The cycle replacement rule (\ref{subrule}) can also be generalized to the
constant field case.

The master formula (\ref{scalarqedmasterF}) is valid off-shell,
and can therefore be used to construct the quenched (one-electron-loop)
Euler-Heisenberg Lagrangians in scalar or spinor QED at the $n$ - loop order
by starting from the one-loop $2(n-1)$ photon amplitude in a constant field,
and sewing off pairs of legs. (Alternatively, one can use the worldline formalism 
also to calculate the  Euler-Heisenberg Lagrangians directly at the multiloop level
\cite{merescsc,mefrss,mereview}.)

\section{The one-loop Euler-Heisenberg Lagrangians}
\label{oneloop}
\renewcommand{\theequation}{3.\arabic{equation}}
\setcounter{equation}{0}

The one-loop EHL's involve only the determinant factors 
in (\ref{calGBGFdet}).
After renormalization, one has

\begin{eqnarray}
{\cal L}_{\rm scal}^{(1)}(F)&=&  {1\over 16\pi^2}
\int_0^{\infty}{dT\over T^3}
\,{\rm e}^{-m^2T}
\biggl[
{(eaT)(ebT)\over \sinh(eaT)\sin(ebT)} 
+{e^2\over 6}(a^2-b^2)T^2 -1
\biggr]
\nonumber\\
{\cal L}_{\rm spin}^{(1)}(F)&=& - {1\over 8\pi^2}
\int_0^{\infty}{dT\over T^3}
\,{\rm e}^{-m^2T}
\biggl\lbrack
{(eaT)(ebT)\over {\rm tanh}(eaT)\tan(ebT)} 
- {e^2\over 3}(a^2-b^2)T^2 -1
\biggr\rbrack\nonumber\\
\label{eh1loop}
\end{eqnarray}
Here  $a,b$  are the two
invariants of the Maxwell field, 
related to $\bf E$, $\bf B$ by $a^2-b^2 = B^2-E^2,\quad ab = {\bf E}\cdot {\bf B}$.
The subtraction terms in the square brackets implement the renormalization of
charge and vacuum energy. The subscripts distinguish between
Scalar and Spinor QED, the superscripts refer to the loop order.

The EHL's contain the information on 
the $N$ photon amplitudes in the low energy limit where all photon energies are small
compared to the electron mass, $\omega_i\ll m$.   At the one-loop four photon
level, this construction of the low-energy amplitude from the effective Lagrangian
is a textbook exercise (see, e.g., \refcite{itzzubbook}).  Even the one-loop (on-shell)
$N$ - photon amplitudes in this limit can still be written quite concisely using
spinor helicity techniques \cite{memascvi}. 

Except for the purely magnetic field case, the parameter integrals in (\ref{eh1loop})
contain poles, leading to an imaginary part of the EHL's. 
A simple application of the residue theorem gives Schwinger's famous
formulas \cite{schwinger51},

\begin{eqnarray}
{\rm Im} {\cal L}_{\rm spin}^{(1)}(E) &=&  \frac{m^4}{8\pi^3}
\beta^2\, \sum_{k=1}^\infty \frac{1}{k^2}
\,\exp\left[-\frac{\pi k}{\beta}\right]
\nonumber\\
{\rm Im}{\cal L}_{\rm scal}^{(1)}(E) 
&=&
\frac{m^4}{16\pi^3}
\beta^2\, \sum_{k=1}^\infty \frac{(-1)^{k+1}}{k^2}
\,\exp\left[-\frac{\pi k}{\beta}\right]
\nonumber\\
\label{schwinger}
\end{eqnarray}
($\beta = eE/m^2$). The $k$th term in these sums is interpreted as 
representing the instability of the vacuum with respect to the coherent production of
$k$ electron-positron (reps. scalar-antiscalar) pairs by the field 
(vacuum tunneling). In the following we will concentrate on the weak field limit 
$\beta \ll 1$ where only the $k=1$ term is  relevant.

\section{The two-loop Euler-Heisenberg Lagrangian}
\renewcommand{\theequation}{4.\arabic{equation}}
\setcounter{equation}{0}

The two-loop EHL, involving one internal photon exchange in the loop, was
first calculated by V.I. Ritus, both for Spinor \cite{ritusspin} and Scalar QED  \cite{ritusscal}.
This resulted in a type of rather intractable two-parameter integrals, on which also later
recalculations were not able to substantially improve \cite{ditreubook,merescsc,mefrss}.
However, the first few coefficients of the weak-field expansions of the two-loop
EHL's have been computed \cite{merescsc,medunsch1,medhrs},  
and there are simple closed-form expressions for the case of a (euclidean) self-dual 
field \cite{medunschSD1}. Those allow one to extend the one-loop calculation of the
on-shell low energy $N$ photon amplitudes, mentioned above, to the two-loop level
for the case where all the photon helicities are the same \cite{medunschSD1}. 

As to the imaginary parts, the Schwinger formulas (\ref{schwinger}) generalize
to the two-loop level as follows \cite{lebrit}:

\begin{eqnarray}
{\rm Im} {\cal L}_{\rm spin}^{(2)} (E) &=&  \frac{m^4}{8\pi^3}
\beta^2\,
\sum_{k=1}^\infty
\alpha\pi K_k^{\rm spin}(\beta)
\,\exp\left[-\frac{\pi k}{\beta}\right]
\nonumber\\
{\rm Im} {\cal L}_{\rm scal}^{(2)} (E) &=&  \frac{m^4}{16\pi^3}
\beta^2\,
\sum_{k=1}^\infty
(-1)^{k+1}
\alpha\pi K_k^{\rm scal}(\beta)
\,\exp\left[-\frac{\pi k}{\beta}\right]
\nonumber\\
\label{ImL2}
\end{eqnarray}
($\alpha=\frac{e^2}{4\pi}$) where

\begin{eqnarray}
K_k^{\rm scal,spin}(\beta) &=& -{c_k\over \sqrt{\beta}} + 1 + {\rm O}(\sqrt{\beta})
\nonumber\\
c_1 = 0,\quad && \quad
c_k = {1\over 2\sqrt{k}}
\sum_{l=1}^{k-1} {1\over \sqrt{l(k-l)}},
\quad k \geq 2
\label{expK}
\end{eqnarray}
Thus at two-loop the $k$th Schwinger-exponential appears with a prefactor which
is still a function of the field strength, and of which presently only the lowest order terms in 
the weak-field expansion are known. Still, things become very simple at leading order
in this expansion: Adding the one-loop and two-loop EHL's, one finds, 
e.g. for the spinor QED case \cite{lebrit},
  
\begin{eqnarray}
{\rm Im} {\cal L}_{\rm spin}^{(1)} (E) +
{\rm Im}{\cal L}_{\rm spin}^{(2)} (E) 
\,\,\,\, {\stackrel{\beta\to 0}{\sim}} \,\,\,\,
 \frac{m^4\beta^2}{8\pi^3}
\bigl(1+\alpha\pi\bigr)
\,{\rm e}^{-{\pi\over\beta}}
\label{Im1plus2}
\end{eqnarray}
and this result is spin-independent (but for the normalization). 
In \refcite{lebrit} it was further noted that, if one assumes that in this 
weak-field approximation higher order corrections just lead to an exponentiation,

 \begin{eqnarray}
 \sum_{l=1}^{\infty}
{\rm Im} {\cal L}^{(l)} (E) \,\,\,\, {\stackrel{\beta\to 0}{\sim}} \,\,\,\,
{\rm Im} {\cal L}^{(1)} (E) 
\,{\rm e}^{\alpha\pi}
\label{Imall}
\end{eqnarray} 
then the result  can, in the tunneling picture, be related to the
fact that a created pair gets born at a finite distance, and thus with a negative
Coulomb binding energy. 

\section{An all-loop conjecture from worldline instantons}
\renewcommand{\theequation}{5.\arabic{equation}}
\setcounter{equation}{0}

Already in earlier work (but unknown to the authors of \refcite{lebrit} at the time)
Affleck et al. \cite{afalma} had invented the concept of worldline instantons,
and their principal application was precisely to demonstrate the exponentiation
property (\ref{Imall}), although not for spinor QED but for the scalar QED case. 
Even though neither derivation of (\ref{Imall}) (to be called ``AAM conjecture'' in the following) can be considered rigorous, the fact that it was obtained 
by two very different lines of reasoning makes us confident about its correctness. 
Assuming this to be the case, (\ref{Imall}) is a highly remarkable result, since,
despite of the simplicity of the derivation by \refcite{afalma} in terms of a single 
semi-classical instanton trajectory, it is a true all-loop result, receiving contributions from an infinite set 
of Feynman diagrams (although it is important for the following that only quenched (one fermion-loop)
diagrams contribute in this limit).
Moreover, \refcite{afalma} argue that (\ref{Imall}) is written in terms of the physically
renormalized mass, that is, the counter diagrams for mass renormalization, 
which normally are necessary starting from the two-loop level,
have been taken care of implicitly. This again cannot be considered as rigorously proven,
however what can be shown easily is that, {\sl if} (\ref{Imall}) holds, then the mass
appearing in it must be the physically renormalized one.  Namely, using a Borel dispersion
relation one can show \cite{medunsch1} the following general relation between the prefactor of the first
Schwinger exponential and the leading asymptotic growth of the weak field
expansion coefficients at fixed loop order $l$: Defining these coefficients by

\bear
{\cal L}^{(l)}(E) = \sum_{n=2}^{\infty} c^{(l)}(n) \Bigl(\frac{eE}{m^2}\Bigr)^{2n}
\label{defcnl}
\ear
one has

\bear
c^{(l)}(n) & {\stackrel{n\to \infty}{\sim}} & c^{(l)}_{\infty} \Gamma[2n - 2] \non\\
 {\rm Im}{\cal L}^{(l)}(E) & \sim & c^{(l)}_{\infty}\,{\rm e}^{-\frac{\pi m^2}{eE}} \non\\
 \label{clnf}
\ear
with constants $c^{(l)}_{\infty}$. This implies, in particular, that the leading factorial
growth order of the expansion coefficients must be the same at each loop order, and
in \refcite{medunsch1} it has been shown that already at the two-loop level this holds true
if and only if the renormalized mass is the on-shell one. 

\section{Relation to the multiloop $N$ photon amplitudes at large $N$}
\renewcommand{\theequation}{6.\arabic{equation}}
\setcounter{equation}{0}

Now, the AAM conjecture is remarkable not only for its simplicity, but also for the fact that, despite of
arising from a true all-order loop summation, the result is analytic in $\alpha$. 
This appears to run contrary to many arguments which have been given, starting with Dyson's classic 1952 paper \cite{dyson52}, to show that S-Matrix elements in QED can never be analytic in $\alpha$.
Now the Schwinger pair creation rate is not itself an S-matrix element, but it can, at any loop order,
be related to the $N$ - photon amplitudes at large $N$ using the
above relations (\ref{defcnl}),(\ref{clnf}) and the already mentioned standard procedure 
for converting the weak field expansion coefficients into
low-energy photon amplitudes (this involves also an extension of the AAM conjecture from
the electric field case to the general constant field or at least self-dual field case \cite{medunschSD1,medunschSD2},  as well as other modest assumptions). This led one of the authors and G.V. Dunne to conjecture in 2004 \cite{mecolima} that the perturbation series for the $N$ photon amplitudes, albeit divergent for the full amplitude, is indeed convergent at the level
of the quenched approximation. This conjecture is not in contradiction with existing general theorems
on the QED perturbation series, and extends a corresponding conjecture made by
Cvitanovic in 1977 \cite{cvitanovic1977} for the $g-2$ factor. 
If true, it would indicate extensive cancellations between
Feynman diagrams, presumably due to gauge invariance.

Referring for the details to \cite{mecolima,medunschSD1,medunschSD2,mehumcsc}, let
us just state here that this line of reasoning leads to a number of nontrivial
predictions starting from the three-loop level. Namely,
we expect to find the expansion coefficients of the three-loop EHL  (for both Scalar and Spinor QED) 
to have the following three properties:

\begin{enumerate}

\item
$\qquad \lim_{n\to\infty} \frac{c^{(3)}(n)}{c^{(1)}(n)} = \frac{1}{2} \alpha^2$.

\item
Only the quenched part of the EHL should contribute to this limit.

\item
The convergence of 
${\frac{c^{(3)}(n)}{c^{(1)}(n)}}$ to ${\frac{c^{(3)}_{\infty}}{c^{(1)}_{\infty}}}$
(from the first eq. in (\ref{clnf})) should not be slower than the one for the corresponding
two-loop to one-loop ratio.

\end{enumerate}

Unfortunately, a calculation of the three-loop EHL has so far been proven technically
out of reach. However, motivated by work of Dunne and Krasnansky \cite{dunkra,krasnansky} on EHL's
in various space-time dimensions it was shown in \refcite{mehumcsc} that the whole
above machinery can, {\sl mutatis mutandis}, be transferred to the computationally
simpler context of QED in 1+1 dimensions. In particular,  all of the above three-loop
predictions can be generalized to this case, changing only the definition of the
ratios,

\bear
\frac{c^{(l)}(n)}{c^{(1)}(n)}  \to  \frac{c_{2D}^{(l)}(n)}{c_{2D}^{(1)}(n+l-1)} 
\label{changeratios}
\ear
and changing $\alpha$ to $2\frac{e^2\pi}{m^2}$.
Preliminary results on a calculation of the three-loop EHL in 2D Spinor QED
using the Feynman diagram approach 
were presented at the QFEXT09 conference \cite{meqfext09},  however this
approach ultimately failed, since it led to parameter integrals with spurious IR divergences.
Here, we will present the results of a new run on the calculation of the same
three-loop EHL using the worldline formalism \cite{in prep}, which has allowed
us to obtain this Lagrangian in terms of manifestly (IR and UV) finite
parameter integrals.

\section{The  three-loop Euler-Heisenberg Lagrangian for 2D QED}
\renewcommand{\theequation}{7.\arabic{equation}}
\setcounter{equation}{0}

At the three-loop level, there are three Feynman diagrams contributing to the
EHL in spinor QED, shown in fig. 2: 

\begin{figure}[h]
{\centering
\includegraphics{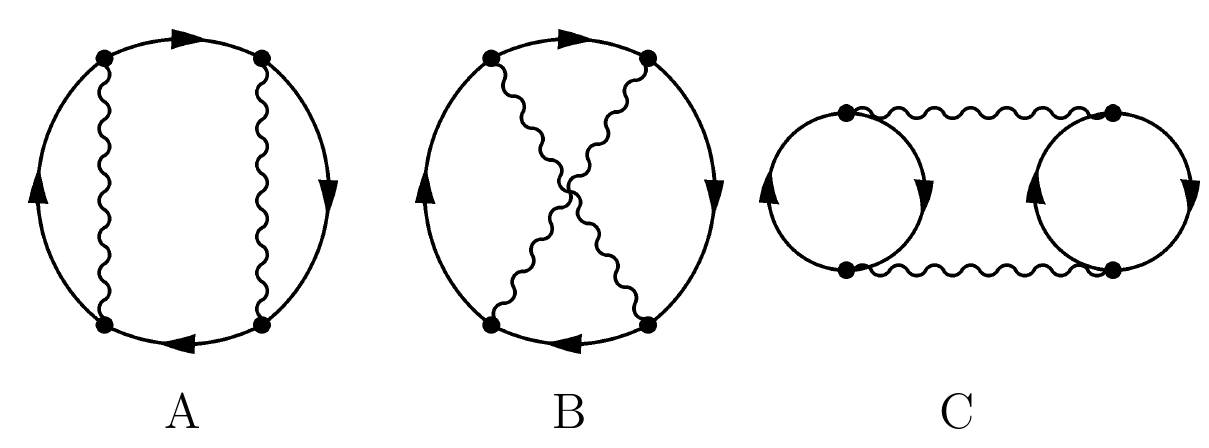}
}
\caption{Feynman diagram representation of the three-loop EHL.}
\label{figEHLexpand}
\end{figure}
\bigskip
Here the solid line stands for the fermion propagator in the constant field with field strength
tensor 
$F=\bigl( \begin{smallmatrix} 
  0 & f\\
  -f & 0
\end{smallmatrix} \bigr)$.
The calculation of the non-quenched diagram C is straightforward, and leads
to a fourfold proper-time integral

\begin{eqnarray}
\mathcal{L}^{3C}(f) &=& \frac{e^3}{16\pi^3 f}\int_{0}^{\infty} dz dz' d\hat{z} dz'' \frac{\sinh z \sinh z' \sinh \hat{z} \sinh z''}{[\sinh (z+z') \sinh (\hat{z}+z'')]^2}\nonumber  \\
& \times & \frac{{\rm e}^{-2\kappa (z+z'+\hat{z}+z'')}}{\sinh z \sinh z' \sinh (\hat{z}+z'') +\sinh \hat{z} \sinh z'' \sinh (z+z')} \nonumber\\
\label{diagC}
\end{eqnarray}
where $\kappa = m^2/2ef$.
From this we have obtained, by numerical integration using MATHEMATICA, the first 12 weak field expansion coefficients, which was sufficient
to verify that they are indeed asymptotically suppressed, even exponentially,
with respect to the asymptotic prediction of the (2D analogue of) the AAM formula. 
Thus point (2) of our three-loop predictions above has been settled.

To the contrary, the parameter integral representations which we have obtained 
for the quenched diagrams $A$ and $B$ are lengthy, and cannot be given here in full.
Their structure is

\begin{eqnarray}
{\cal L}^{3(A+B)}(f)
&=&
-\frac{e^4}{(4\pi)^3}
{ \int_{0}^{\infty}}{dT\over T^2}
{\rm e}^{-m^2T}
\frac{Z}{\tanh Z}
\prod_{i=1}^4 \int_0^T 
d\tau_i
\nonumber\\&&\times
\Bigl(2I_{1234}+I_{1324}+4I_{123}+2I_{12}+4I_{13}+I_{12,34}+2I_{13,24}\Bigr)
\nonumber\\
\label{AB}
\end{eqnarray}
where $Z=efT$ and, for example,

\begin{eqnarray}
I_{ijkl} &=& \frac{{\rm tr}\bigl(\lbrace ijkl\rbrace_S\bigr)}{\Delta} \label{defIijkl}
\end{eqnarray}
with (compare (\ref{subrule}))

\bear
\lbrace i_1i_2\ldots i_n \rbrace_S &:=&
\Gphat{i_1}{i_2}\Gphat{i_2}{i_3}\cdots\Gphat{i_n}{i_1} 
-
{\cal G}_{Fi_1i_2}{\cal G}_{Fi_2i_3} \cdots {\cal G}_{Fi_ni_1}
\label{defbrace}
\ear
Here we have further introduced 
$\hat{\dot{\mathcal{G}}}_{Bij}:=\dot{\mathcal{G}}_{Bij}-\dot{\mathcal{G}}_{Bii}+\mathcal{G}_{Fii}$,
and  $\Delta$ is a determinant also involving the worldline Green's functions.
Note that (\ref{AB}) represents the sum of both diagrams $A$ and $B$.

\section{Conclusions}
\renewcommand{\theequation}{8.\arabic{equation}}
\setcounter{equation}{0}

We have summarized here what is presently known about multiloop corrections to the
EHL, concentrating on the potential of such corrections to yield information on the
high-order behavior of the perturbation series in QED. As part of a long-term effort to
prove or disprove ``quenched convergence'' 
for the case of the photon S-matrix, we have presented a
parameter integral representation of the three-loop EHL in 2D QED. Although the
extraction of the weak-field expansion coefficients from this representation and 
verification of the remaining predictions implied by this conjecture at the three-loop level
(points (1) and (3) above)
will still require very substantial work, we are confident that we will have definite results to show by
the time of QFEXT13!

\section*{Acknowledgments}

C.S. and I.H. thank CONACYT for financial support.

\end{document}